# DC high voltage to drive helium plasma jet comprised of repetitive streamer breakdowns

Xingxing Wang, Alexey Shashurin

Purdue University, School of Aeronautics and Astronautics, West Lafayette, IN  47907, USA

**Abstract**

This paper demonstrates and studies helium atmospheric pressure plasma jet comprised of series of repetitive streamer breakdowns, which is driven by a pure DC high voltage (auto-oscillations). Repetition frequency of the breakdowns is governed by the geometry of discharge electrodes/surroundings and gas flow rate. Each next streamer is initiated when the electric field on the anode tip recovers after the previous breakdown and reaches the breakdown threshold value of about 2.5 kV/cm. Repetition frequency of the streamer breakdowns excited using this principle can be simply tuned by reconfiguring the discharge electrode geometry. This custom-designed type of the helium plasma jet, which operates on the DC high voltage and is comprised of the series of the repetitive streamer breakdowns at frequency about 13 kHz, is demonstrated.

Non-equilibrium atmospheric plasma jets (NEAPJ) are widely used nowadays in the fields of bio-engineering, medicine, food processing etc. [1]˒ [2]˒ [3]˒ [4]˒ [5] Direct sterilization by NEAPJ of open wounds, ulcers or burns is more efficient compared to ordinary sterilization using chemicals due to the potential extend damage that chemicals may cause to punctured tissues and organs. [6] NEAPJ can also be used for treatment of various types of cancer including lung, bladder, skin, head and neck, brain, pancreatic tumors etc. [2]˒ [3]˒ [6]˒ [7]˒ [8]

Conventional NEAPJ are excited in helium (He) flow exhausted from the discharge tube into open air. Multiple parameters of He plasma jets were measured previously including plasma density, temperatures of various species, electrical currents etc. [9] Typically, plasma electron density $n_e$ is in the range of $10^{12}$-$10^{13}$ cm$^{-3}$ while the temperature of heavy species is near the room temperature at 300-350 K. [10]

Conventional NEAPJ are excited using AC or pulsed DC power supplies operating in kV range and frequencies around 10s of kHz. [10]˒ [11]˒ [12]˒ [13]˒ [14]˒ [15]˒ [16] In those cases, breakdown takes place once every cycle of the applied high voltage (HV) when the voltage applied to the electrode reaches the breakdown threshold. The breakdown is associated with development of streamer tip propagating at characteristic velocities in the range of $10^6$ -$10^8$ cm/s increasing with the high voltage magnitude. [10] The duration of each individual streamer does not exceed the period of time of several µs and stops where the presence of the oxygen in the He jet increases along the jet to about 1 percent. [10]˒ [17]˒ [18]  The plasma remaining in the streamer channel decays shortly afterwards (on time scale of about several µs). The next breakdown event occurs on the next cycle of the applied AC high voltage or with the application of the next high voltage pulse. Thus, the repetition frequency of the discharge is fully governed by the discharge driving power supply operation frequency. [16]

Therefore, excitation of the helium NEAPJ by AC and pulsed DC high voltage was studied previously. However, it is interesting to evaluate excitation of the helium NEAPJ by means of constant DC high voltage applied to the electrode. Obviously, simultaneously with the initial application of the DC high voltage the first streamer will be initiated similarly to that in DC pulsed excitation. Shortly after the firing, the streamer stops and plasma column decays (about few µs after the firing). [10] [19] [20] [21].  Once plasma decay is complete, the system returns back to the conditions which, from the first glance, are indistinguishable from those existing prior to the shooting of the initial streamer. Then, obvious questions arise: Will the second streamer be fired in the system and what governs the timing of its firing? How can the repetition frequency of these streamers be controlled?

This paper studies the He plasma jet comprised of a series of repetitive streamer breakdowns which is excited by pure DC high voltage and demonstrates the ways to control the frequency of streamer repetition. Cold plasma gun operating on this principle is presented.

The schematics of the circuitry are shown as Figure 1. A DC power supply up to 5 kV was used in the experiments. A mass flow controller (Sierra SmartTrak 100) was used to control the He supply into the nozzle. The high-voltage electrode on the axis of the nozzle was immersed into the He which flowed through the nozzle into open air. The inner diameter of the nozzle exit was 3.6 mm. A grounded metal sheet was installed outside the nozzle at the distance $d$. The electrical current was measured using a 10 k$\Omega$ shunt resistor placed in series in the HV line as shown in Figure 1.

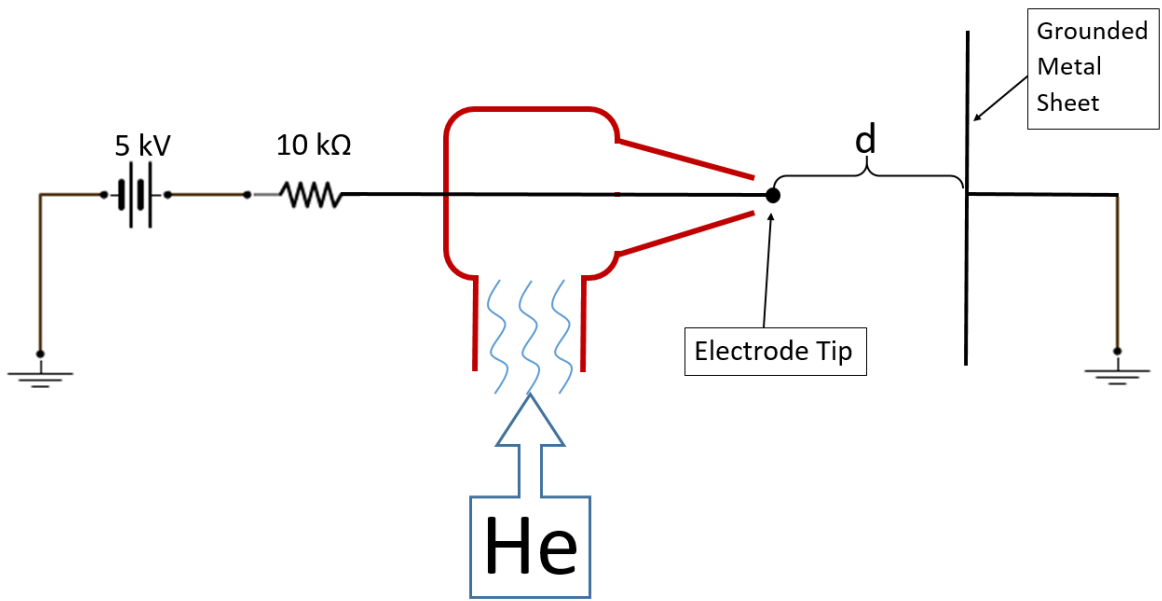

**Figure 1 Schematics of the experiment setup**

Series of the streamer breakdowns were observed when 5 kV DC voltage was applied to the electrode as shown in Figure 2 (a). The time interval between two adjacent streamer breakdown pulses was 400-750 μs and amplitude of the current pulse was about 0.15-0.25 mA. A close look of the current and voltage waveforms of an individual discharge is shown in Figure 2 (b). It is important to note that the applied HV remains constant throughout the entire duration of the experiment including the period of streamer breakdown. In this setup the ground metal sheet was placed far away from the nozzle in order to eliminate its effect on the discharge behavior, whereas the surroundings were considered to provide the ground potential for discharge. Current pulse width was about 1.5 µs. The streamer development was photographed using an intensified charge-coupled device (ICCD) camera as shown in Figure 2 (c) for the moments of time *[t1]-[t6]* indicated by the rectangular bars shown in Figure 2 (b). Average velocity of streamer front propagation was ~$2.5·10^6$ cm/s. Note, repetitive streamer breakdowns presented in Figure 2 are different from self-pulsing atmospheric-pressure plasma jets studied recently [22], [23], [24], [25]. Even though DC high voltage sources were utilized in these works, measurement of voltage and current waveforms revealed dramatic reduction of the electrode voltage every time breakdown occurred, similar to fairly well-studied pulsed DC atmospheric-pressure plasma jets. [10]

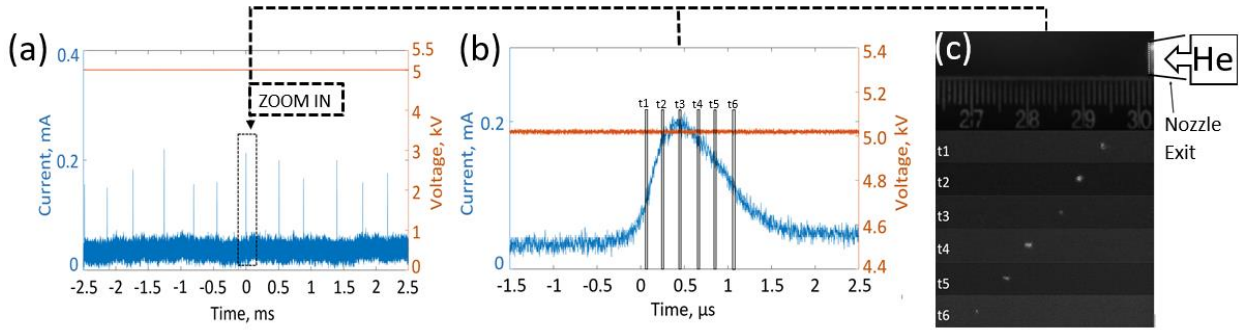

**Figure 2 (a) Current and voltage waveforms of series of the streamer breakdowns excited by the 5 kV DC voltage supplied to the electrode. (b) Current and voltage waveforms of an individual streamer breakdown. *[t1]-[t6]* are the time periods when a photograph of the streamer was taken. (c) Photographs of the streamer at certain time from *[t1]* to *[t6]*. The grounded electrode was located at d ≈ 0.5 m. The He flow rate was 1 L/min.**

The phenomenon of repetitive streamer pulsations is known quite well for streamer corona developing in air and called flashing streamer corona. [26] The frequency of repetitive pulsations of the DC driven streamer corona in air is typically within around 5-10 kHz. [26], [27] For example, the repetitive streamer breakdowns were observed at the frequency < 6.5 kHz when 5 – 9.3 kV DC voltage was applied to a needle electrode of radius 0.17 mm placed 31 mm away from the cathode plate. [26] The repetition frequency is governed by the following process. During each streamer breakdown, electrons which are produced by ionization of air on the streamer path are flowing into the anode electrode. This process can be easily understood by imagining "metallization" of the regions occupied by the streamer channel. The term "metallization" is used here to indicate presence of finite electrical conductivity of plasma along the streamer channel as oppose to the non-conducting air. For example, for ideally conducting streamer channel positive potential of the anode will be establishing along the entire streamer path (of course, in reality the potential along the streamer channel drops below the anode potential due to finite conductivity of plasma in the streamer channel). Establishing of this positive potential along the streamer path causes removal the excess of the electron charge from the streamer channel to the anode electrode and build-up of the positive space charge along the streamer path. Later, when the streamer growth stops and the plasma in the streamer channel decays, the remaining positive space charge compensates the electric field near the anode, which prevents initiation of the new streamer. However, as far as the positive space charge drifts toward the cathode, the electric field near the anode is restored. New streamer is initiated when the near-anode electric field is reaching the threshold value required for the breakdown.

Therefore, it might be concluded that initiation of each next streamer breakdown occurs when the electric field at the anode recovers after the previous breakdown and exceeds the threshold value $E_{th}$. Moreover, the value of $E_{th}$ is universal for the fixed gas composition and the pressure. Indeed, geometry of electrodes or other parameters of the system can cause redistribution of the electric field, but cannot affect the $E_{th}$ value. The streamer will fire at the location where $E_{th}$ is reached first. The value of $E_{th}$ was determined experimentally as follows.

Voltage was applied to the spherical electrode immersed in the He flow using the setup shown in Figure 1. The voltage was increased from zero up to the value $U_{th}$, when firing of the first streamer was detected (no space charge remaining from the previous breakdowns). In this case, the electric field ($E$) in vicinity of the spherical high voltage electrode is related to the tip voltage ($U$) as $E = \frac{U}{a}$, where $a$ is the radius of the electrode tip sphere. Thus, the threshold electric field ($E_{th}$) was determined as:

$$E_{th} = \frac{U_{th}}{a}. \qquad (1)$$

Note, the threshold electric field $E_{th}$ is defined as the electric field on the electrode prior to the streamer initiation and thus, governed by the electrode radius $a$ as shown in Eq. (1), [26] while local electric field around the streamer tip during the following streamer growth is governed by the radius of the streamer channel ($r_s$) and can be written as $\frac{U}{2r_s}$ for ideally conductive streamer channel. [28]

In order to evaluate critical electric field $E_{th}$ required to fire the streamer, we used three spherical HV electrodes of diameters 1.59 mm, 2.38 mm and 3.18mm. The dependence of $U_{th}$ required to fire the first streamer on electrode tip diameter is shown in Figure 3. One can see that the dependence was linear and crosses the origin. This clearly indicates that threshold electric field strength was constant, $E_{th}$ = 2.5 kV/cm, for all cases, which supports the predictions formulated above.

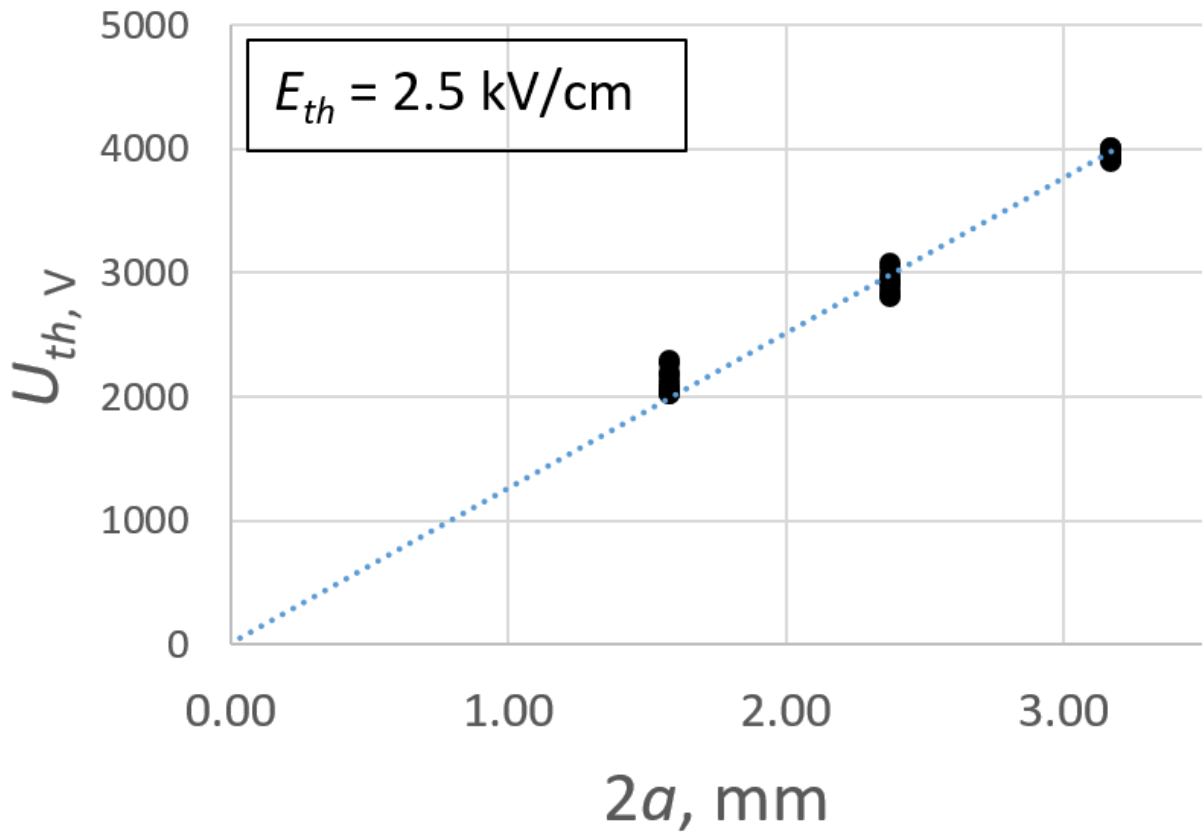

**Figure 3 Breakdown voltage required for triggering the first streamer for three different sizes of spherical electrodes. Threshold electric field required for the breakdown is about 2.5 kV/cm regardless the electrode size.**

Let us now consider the ways how frequency of repetition of streamer breakdowns can be controlled. Figure 4 shows the time interval between two adjacent streamer breakdowns ($T_{rep}$) vs. He flow rate. One can see that the time interval between breakdowns was reduced with the increase of He flow rate. This can be explained by the fact that larger He flow rate leads to faster He flow speed and this in turn results in faster removal of the positive space charge from the vicinity of electrode tip. Thus, electric field in vicinity of the streamer recovered faster to the threshold value $E_{th}$ which led to more frequent pulsations. In addition, it was observed that more frequent streamer pulsations were more repeatable which is seen by decrease of the error bar for higher He flow rates.

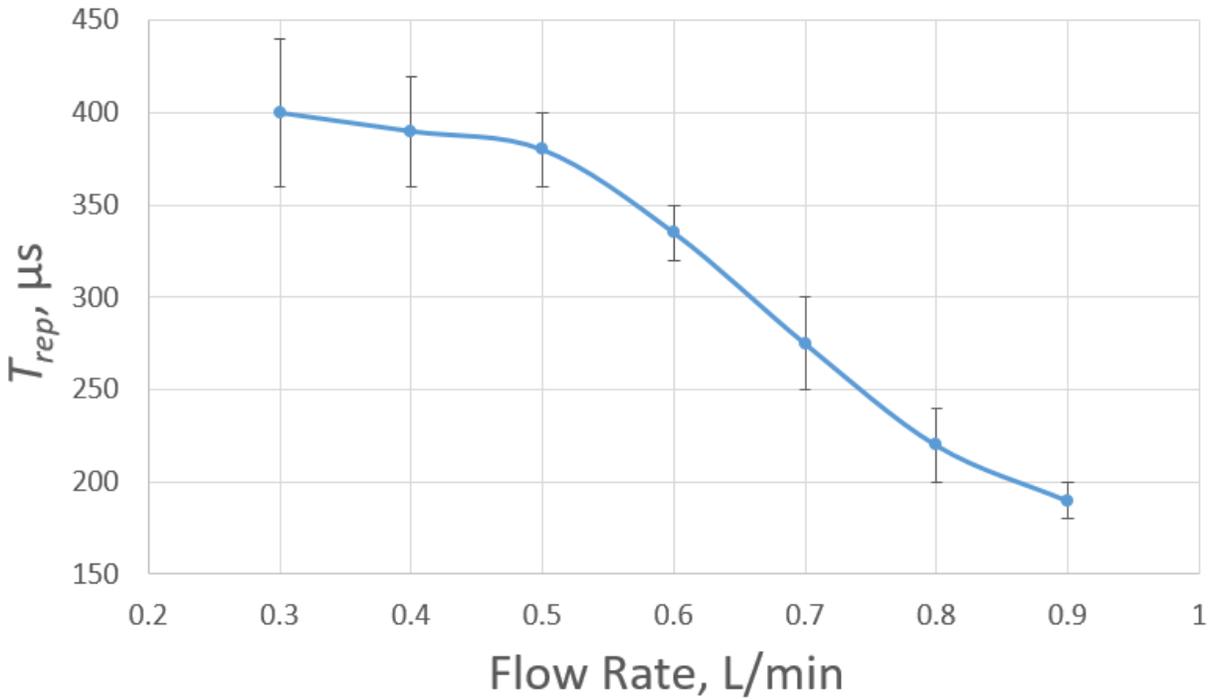

**Figure 4 Streamer breakdowns vs. He flow rate for *d* = 5 cm and *U* = 5000 V.**

Another way to control the repetition time interval of streamer firing was to vary the background potential around the tip electrode. Indeed, electric field near the tip of spherical electrode at potential $U$ immersed in space with background potential $U_0$ can be written as [28]: $E = \frac{U-U_0}{a}$. The value of background potential can be decreased $U_0 < 0$, for example, by decreasing distance $d$ between the HV electrode tip and grounded metal plate (see Figure 1), due to the contribution of 'negative mirror-image charge' behind the plate according to the method of images in electrostatics. [29] The experimental confirmation of this fact is demonstrated in Figure 5 showing the relation between the repetition time interval and the distance *d*. One can see from the plot that the decrease of the distance *d* caused more frequent streamer firings (period between the streamer firings $T_{rep}$ decreased). More frequent streamer firings were associated with more stable and repeatable operation as indicated by the size of error bars in Figure 5.

It has to be noted, the experimental fact that $T_{rep}$ increases for smaller *d* (Figure 4) and for larger flow rate (Figure 5) was obtained and confirmed multiple times in single experimental runs consisting in real-time manual adjustment of distance *d* (or flow rate) in both directions and simultaneous observation of corresponding changes of $T_{rep}$ on oscilloscope. However, spread of the measured $T_{rep}$ between different experimental runs was quite high (up to 30%) due to high sensitivity of the source to precise geometry of grounded surroundings which is extremely hard to reproduce exactly the same in the different experimental runs. This spread can be traced by comparing of two $T_{rep}$ measurements conducted at close conditions, but obtained in different experimental runs namely, $T_{rep}$=190 and 250 μs for *d*=5 cm and very close He flows 0.9 and 1 L/min (see Figure 4 and Figure 5).

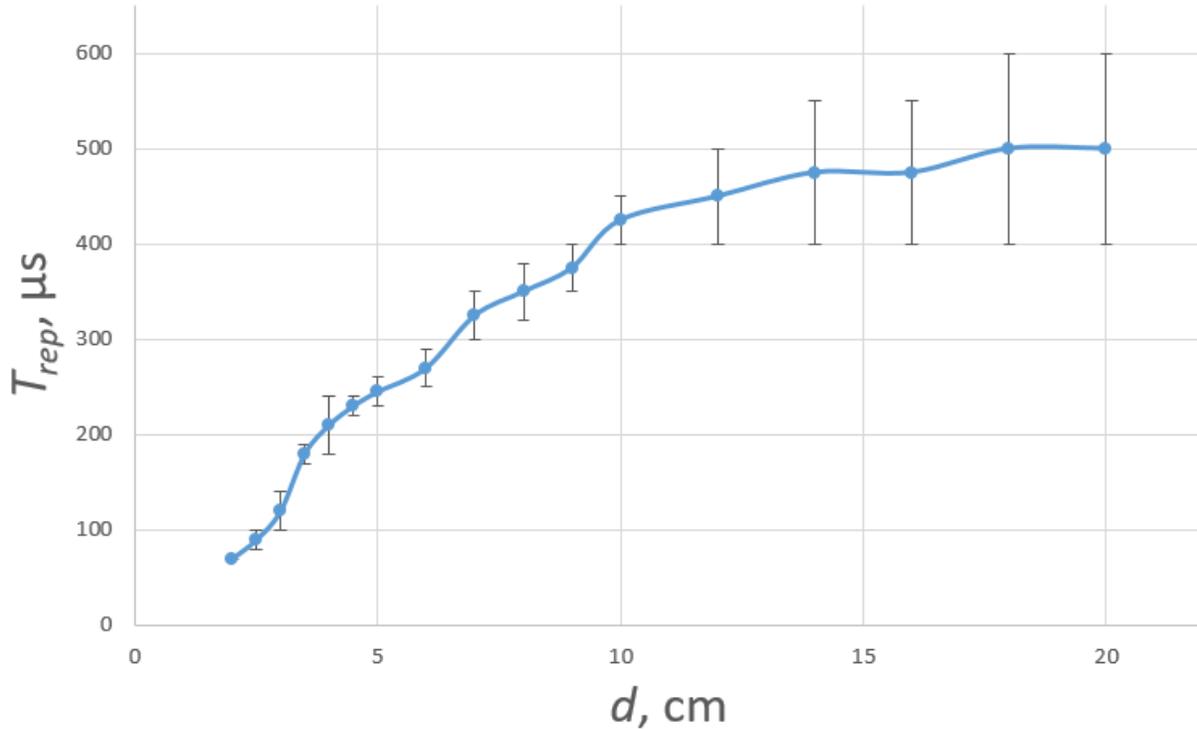

**Figure 5** Average period between streamer breakdowns vs. distance to the grounding plate d. Flow rate = 1 L/min, *U*= 5000 VDC.

It is interesting that the geometrical control of the repetition frequency of the streamer breakdowns can be utilized for creating another type of the helium NEAPJ as follows. The design shown in Figure 6 uses a customized glass pipette and a funnel attached to it. The high voltage electrode is inserted into the tube near the nozzle exit while grounded metal ring is fixed on the funnel. The grounded ring was used in order to vary background potential and thus to adjust the repetition frequency of the streamer breakdowns. Helium flow went through the center of the symmetric grounded ring. Several ring sizes and the distances from the pipette exit were tested in order to achieve steady and visible plasma as well as to extend the length of plasma. Within this range, the location of the grounded ring was determined in the design shown in Figure 6 in order to fix the repetition frequency at around 15 kHz, which is typical for conventional NEAPJ plasma guns. The length of the free plasma jet was approximately 1 cm, while it can be extended to more than 2 cm if a finger is placed in vicinity as shown in in Figure 6 (b). The voltage and current during the streamer breakdown are shown in Figure 7. One can see that current amplitude was about 0.1 mA and repetition frequency - 13 kHz. NEAPJ plasma gun presented in Figure 6 is a dielectric barrier discharge with central electrode in contact with plasmas and grounded ring electrode is insulated from the direct contact with plasmas. By comparing Figure 2 and Figure 7, one can see that the duration of the discharge increases by approximately factor of 2. This can be explained by the closer proximity of the ground electrode for the case shown in Figure 7 that slows down the streamer front propagation and extend its temporal duration. [30]

In has to be noted that the main difference of the considered here phenomenon of repetitive streamer firing compared to the classical flashing streamer corona in uniform air is that the firing of the streamer occurs strictly along the He jet exhausting from the discharge tube to an ambient air. The streamer propagation is confined to the He jet due to elevated electron attachment to oxygen molecules in direction perpendicular to the He flow. [10] This predetermined streamer path as oppose to the random path of the air steamer and, therefore, enables unique opportunities for geometric control of streamer firing frequency demonstrated in this work.

In conclusion, we have demonstrated mechanism of excitation of helium plasma jet comprised of repetitive streamer breakdowns which is driven by a purely DC high voltage. This type of the plasma source allows control of operation frequency simply by varying the geometry of the discharge electrodes. Another type of DC-voltage driven helium plasma jet comprised of pulsed streamer breakdowns is demonstrated.

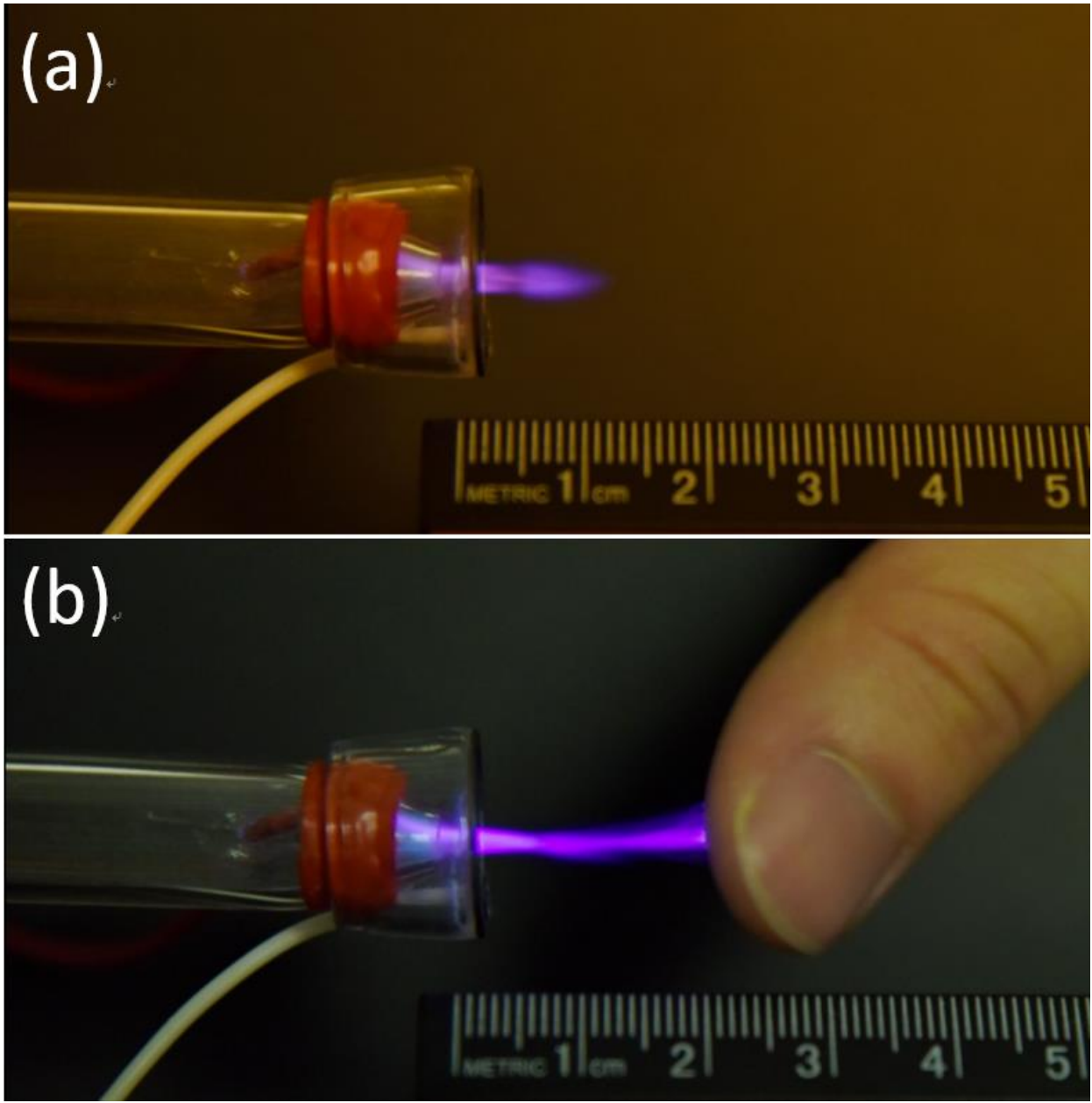

*Figure 6 Photographs of the NEAPJ driven by DC high voltage. (a) Length of the free jet is about 1 cm, (b) Length of the plasma jet can be extended to 2 cm if finger is placed nearby. He flow rate is 2 L/min and U = 3400 V.*

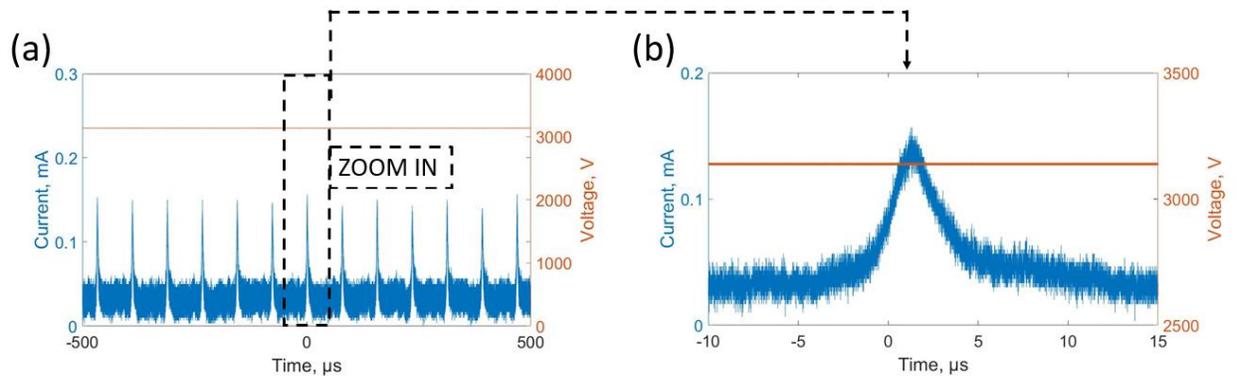

**Figure 7.** Current and voltage waveforms of the NEAPJ driven by DC high voltage. (a) Multiple breakdown events. (b) Temporally resolved individual breakdown.

**Acknowledgments**

Authors would like to thank Dr. Macheret for fruitful discussions.